\documentclass[12pt,fleqn]{article}

\usepackage{graphicx}
\usepackage{pdftricks}

\begin{document}

\title{\Large Unitary thermodynamics from thermodynamic geometry}

\author{
   George Ruppeiner\footnote{ruppeiner@ncf.edu}\\
   Division of Natural Sciences\\
   New College of Florida\\
   5800 Bay Shore Road\\
   Sarasota, Florida 34243-2109}

\maketitle

\begin{abstract}

Degenerate Fermi gases of atoms near a Feshbach resonance show universal thermodynamic properties, which are here calculated with the geometry of thermodynamics, and the thermodynamic curvature $R$.  Unitary thermodynamics is expressed as the solution to a pair of ordinary differential equations, a ''superfluid'' one valid for small entropy per atom $z\equiv S/N  k_B$, and a ''normal'' one valid for high $z$.  These two solutions are joined at a second-order phase transition at $z=z_c$.  Define the internal energy per atom in units of the Fermi energy as $Y=Y(z)$.  For small $z$, $Y(z)=y_0+y_1 z^{\alpha }+y_2 z^{2 \alpha}+\cdots,$ where $\alpha$ is a constant exponent, $y_0$ and $y_1$ are scaling factors, and the series coefficients $y_i$ ($i\ge 2$) are determined uniquely in terms of $(\alpha, y_0, y_1)$.  For large $z$ the solution follows if we also specify $z_c$, with $Y(z)$ diverging as $z^{5/3}$ for high $z$.  The four undetermined parameters $(\alpha,y_0,y_1,z_c)$ were determined by fitting the theory to experimental data taken by a Duke University group on $^6$Li in an optical trap with a Gaussian potential.  The very best fit of this theory to the data had $\alpha=2.1$, $z_c=4.7$, $y_0=0.277$, and $y_1=0.0735$, with $\chi^2=0.95$.  The corresponding Bertsch parameter is $\xi_B=0.462(40)$.

\end{abstract}

\noindent
{\bf Keywords}: unitary thermodynamics; thermodynamic curvature; strongly interacting Fermi systems; Feshbach resonance; ultracold quantum gases

\section{INTRODUCTION}

There has been considerable recent interest in strongly interacting degenerate systems of atomic Fermi gases as models for quark-gluon plasmas, neutron star matter, and high temperature superconductors \cite{Luo2007, Horikoshi2010}.  Such atomic systems have been studied at low temperatures in optical traps with magnetic fields tuned to produce states near Feshbach resonance \cite{Chin2010}.  Low density conditions are produced where the atomic s-wave scattering length has absolute value much greater than the average interatomic spacing, which in turn is much greater than the pair interaction length.  We expect a universal thermodynamics, identical for all systems belonging to such a class of systems \cite{Ho2004}.  I propose to calculate this unitary thermodynamics using thermodynamic methods based on the thermodynamic curvature.

\par
At the outset, it is important to distinguish between three volume regimes: 1) macroscopic volumes in the thermodynamic limit, 2) microscopic volumes at the level of the individual atoms, and 3) mesoscopic volumes at the level of the correlation length $\xi$.  It is at these mesoscopic volumes that significant elements of the system properties get determined.

\par
Mesoscopic volumes at length scale $\xi$ offer both challenges and opportunities.  If $\xi$ were less than the order of the average distance between atoms, then the ability of interatomic interactions to organize the system into interesting mesoscopic structures is weak.  We have then some approximation of the ideal gas, readily dealt with by a number of perturbation schemes in statistical mechanics.  On the other hand, if $\xi$ encompasses a large number of atoms, then computing from the microscopic level up with statistical mechanics can be very difficult.  Special techniques, such as renormalization group theory, may be required.  However, cases with large $\xi$ frequently posses thermodynamic properties independent of the details of interatomic interactions.  Such ''universality'' can lead to simplification.  Methods of exploiting situations with large $\xi$, such as the one in this paper, could be very productive.

\par
I take an entirely thermodynamic approach for determining unitary thermodynamics.  This thermodynamic approach taps into mesoscopic fluctuations on a large scale in an attempt to bring out universal properties.  The calculation is based on solving the differential equation resulting from setting the thermodynamic curvature $R$ proportional to the inverse of the thermodynamic potential per volume \cite{Ruppeiner1995}.  The very difficulty posed by having too many atoms to calculate with in statistical mechanics with large $\xi$ makes the thermodynamic approach effective.  Thermodynamics works not by calculating over individual atoms, but by averaging over many atoms.  Thermodynamic fluctuation theory, including thermodynamic curvature $R$, allows us to work with fluctuating mesoscopic structures with thermodynamic methods.  These ideas get augmented with hyperscaling from the theory of critical phenomena.

\par
The resulting solution for unitary thermodynamics comes in two parts, connected at a critical value $z_c$ of the dimensionless entropy per atom

\begin{equation} z\equiv \frac{S}{ k_B N}, \label{02} \end{equation}

\noindent where $S$ is the entropy, $N$ is the number of atoms in the system, and $k_B$ is Boltzmann's constant.  There is a ''superfluid'' phase for $z<z_c$ and a ''normal'' phase for $z>z_c$.  I will assume that the joining point $z=z_c$ corresponds to a second-order phase transition.

\par
I find that the internal energy $Y$ per atom, in units of the Fermi energy, is a function of just $z$, $Y=Y(z)$.  This scaling principle is standard in these applications \cite{Ho2004}.  For small $z$, I find

\begin{equation} Y(z) = y_0+y_1 \, z^{\alpha }+y_2 \, z^{2 \alpha}+\cdots, \label{05} \end{equation}

\noindent where $\alpha$ is a constant exponent, $y_0$ and $y_1$ are simply related to scaling factors for $z$ and $Y(z)$, and the remaining series coefficients $y_i$ ($i\ge 2$) are determined uniquely in terms of $(\alpha,y_0,y_1)$.  I find that for large $z$, $Y(z)$ goes asymptotically to infinity as $Y(z)\propto z^{5/3}$.

\par
This two-part solution contains four free parameters $(\alpha,y_0,y_1,z_c)$, which may be determined by trap integrating and fitting to experimental data.  I analyzed the data of the Duke University group taken on $^6$Li in an optical trap with a Gaussian potential \cite{Luo2009}.  The fitting procedure puts reasonably stringent constraints on $(\alpha, y_0, y_1)$, but is less restrictive on $z_c$.  The very best fit had $\alpha=2.1$, $z_c=4.7$, $y_0=0.277$, and $y_1=0.0735$, with goodness of fit $\chi^2=0.95$.  The corresponding Bertsch interaction parameter was $\xi_B=0.462(40)$.  But good fits with smaller $z_c$ were also found.  For example, $\alpha=2.1$, $z_c=3.0$, had $\chi^2=1.44$, and $\xi_B=0.434$, in agreement with $\xi_B = 0.435(15)$ determined in the Duke experiment with speed of sound measurements \cite{Luo2009}.  Placing a precise upper limit on $z_c$ was difficult, but fits using only the lower segment for $Y(z)$ were not as good as the two-segment fits.  I conclude that a phase transition is indicated, but hard to locate precisely.

\par
This paper starts with a description of the method of calculating thermodynamic properties from Riemannian geometry of thermodynamics, a discussion which features the thermodynamic curvature $R$.  Second, I present a determination of the two-part solution to the geometric equation.  Third, I present the trap integration of the local densities and the fit to the Duke data.

\par
I add that after my analysis of the Duke data was well under way, data were published by Ku {\it et al.} \cite{Ku2012} involving an experiment on a {\it homogeneous} system displaying unitary thermodynamics.  The advantage of such an experiment is that its analysis does not require trap integration.  The theoretical method here could certainly be employed to analyze the experiment of Ku {\it et al.} but to do so was beyond the scope of this project, and will be deferred to the future.

\section{THE GEOMETRIC EQUATION}

In this section, I summarize the properties of the thermodynamic curvature $R$.  I also present the geometric equation for the thermodynamic properties.  I consider only systems consisting of one type of atoms.

\par
Although we may calculate equivalent results with any choice of independent thermodynamic parameters, it is most natural to analyze the Duke data starting from a Local Density Approximation (LDA) expressed in terms of the fundamental thermodynamic equation $E=E(S,N,V)$, where $E$ is the internal energy, $S$ is the entropy, $N$ is the number of atoms, and $V$ is the volume \cite{Callen}.  Define as well the temperature, chemical potential, and pressure: $\{T,\mu,p\}\equiv\{E_{,S},E_{,N},-E_{,V}\}$, where the comma notation indicates differentiation.  My notation is for a uniform thermodynamic system, with a different notation (introduced later) for the nonuniform trap thermodynamics.  The volume integration over the LDA properties yields the theoretical prediction of the trapped thermodynamics.

\par
Define the thermodynamic entropy information metric $(\Delta\ell)^2$ in terms of the fluctuation probability of an open subsystem with fixed volume $V$ of an infinite reservoir in a reference state ''0'' \cite{Pathria, Landau}:

\begin{equation}\mbox{probability}\propto\mbox{exp}\left[-\frac{V}{2}(\Delta\ell)^2 \right].\label{07}\end{equation}

\noindent $(\Delta\ell)^2$ is an invariant, positive definite quadratic form which in the pair of independent thermodynamic parameters $X^1=S$ and $X^2=N$ may be written as

\begin{equation}(\Delta \ell)^2=g_{\mu\nu}\Delta X^\mu \Delta X^\nu,\label{10} \end{equation}

\noindent where $\Delta X^\alpha\equiv (X^\alpha -X^\alpha_0)$ $(\alpha=1,2)$ denotes the difference between the thermodynamic parameters $X^{\alpha}$ of the subsystem and their values $X^{\alpha}_0$ corresponding to $(\Delta \ell)^2 =0$.  The thermodynamic metric elements

\begin{equation}g_{\alpha\beta} \equiv \frac{1}{k_B T V}\frac{\partial^2 E}{\partial X^\alpha\partial X^\beta}\label{20}\end{equation}

\noindent are evaluated in the state $X^\alpha_0$.

\par
A Riemannian metric naturally induces a curvature on the surface of thermodynamic states parameterized by $(X^1,X^2)$, as described in any book on differential geometry \cite{Laugwitz1965}.  The thermodynamic Riemannian curvature scalar (in the sign convention of Weinberg \cite{Weinberg}, where the 2-sphere has $R<0$) may be written as \cite{Laugwitz1965,Weinberg,Ruppeiner2012}

\begin{equation} \begin{array}{lr} {\displaystyle R= -\frac{1}{\sqrt{g}} \left[ \frac{\partial}{\partial X^1}\left(\frac{g_{12}}{g_{11}\sqrt{g}}\frac{\partial g_{11}}{\partial X^2}-\frac{1}{\sqrt{g}}\frac{\partial g_{22}}{\partial X^1}\right) \right. } \\  \hspace{3.6cm} + {\displaystyle \left. \frac{\partial}{\partial X^2}\left(\frac{2}{\sqrt{g}} \frac{\partial g_{12}}{\partial X^1} -\frac{1}{\sqrt{g}}\frac{\partial g_{11}}{\partial X^2}-\frac{g_{12}}{g_{11}\sqrt{g}}\frac{\partial g_{11}}{\partial X^1}\right)\right],}  \end{array} \label{30}\end{equation}

\noindent where

\begin{equation}g\equiv g_{11}g_{22}-g_{12}^2. \label{35}\end{equation}

\noindent $R$ is an intensive thermodynamic quantity with units of volume per atom.  Although the thermodynamic metric elements change their form on transforming coordinates, the value of $R$ for a given thermodynamic state does not change.  $R$ is thus invariant on changing coordinates, by the rules of Riemannian geometry.  For calculating $R$, the choice of coordinates is one purely of convenience.

\par
Riemannian geometry has a reputation as being difficult, mostly because of its application in the four-dimensional theory of general relativity, with its semidefinite spacetime metric.  In the two-dimensional geometry of this paper, with its positive definite metric, the mathematics is considerably simpler.

\par
$R=0$ for the classical ideal gas, suggesting $R$ is a measure of interatomic interactions \cite{Ruppeiner1979}.  Indeed, explicit calculations in a number of cases strongly suggest that $|R|$ is the correlation volume,

\begin{equation} |R|\sim\xi^3, \label{40}\end{equation}

\noindent where $\xi$ is the correlation length \cite{Ruppeiner1995, Ruppeiner1979, Johnston2003}.  This interpretation is also supported by a covariant thermodynamic fluctuation theory \cite{Ruppeiner1983a, Ruppeiner1983b, Diosi1985}.  $R$ appears to be positive for systems with repulsive interactions and negative for systems with attractive interactions \cite{Ruppeiner2010}.  Janyszek and Mruga{\l}a \cite{Mrugala1990} and Oshima {\it et al.} \cite{Oshima1999} first emphasized the difference in the sign of $R$ between the Fermi ($R>0$) and Bose ($R<0$) ideal gasses.  $R$ for the ideal Fermi gas diverges to positive infinity as the temperature goes to zero.

\par
Adding hyperscaling from the theory of critical phenomena to the thermodynamic geometric picture gives us a way to calculate unitary thermodynamics.  Hyperscaling asserts that the singular part of the thermodynamic potential per volume $\phi$ is proportional to the inverse of the correlation volume \cite{Widom74, Goodstein},

\begin{equation} \phi\sim\xi^{-3}.\label{70}\end{equation}

\noindent Combining this with $|R|\sim\xi^3$ in Eq. (\ref{40}) leads to the geometric equation \cite{Ruppeiner1991}:

\begin{equation} R=-\frac{\kappa}{\phi}, \label{80}\end{equation}

\noindent where $\kappa$ is a dimensionless constant of order unity which the solution process will determine.

\par
This derivation of the geometric equation is somewhat loose and approximate.  I present it mainly to give the reader some motivation of where these ideas come from.  In practical applications, it is the geometric equation Eq. (\ref{80}), in conjunction with a background subtraction, which are important.  Their precise expression is motivated by mathematical consistency, and not by the loose derivation above.  If the reader cares to, he or she may simply regard the geometric equation as a postulate, and dispense with the derivation all together.

\par
Appendix 1 gives a detailed discussion of how $\phi$ is defined:

\begin{equation} \phi=\frac{p}{k_B T}.  \label{90}\end{equation}

\noindent This discussion offers few alternatives to the choice made here.  The choice of background subtraction to get the critical properties is equally limited.  There are two possibilities depending on which of two types of singular points the solution is built around: 1) a singular point $\mathcal{P}_0$ with $|R|\to\infty$ for which

\begin{equation} R=-\kappa\left(\frac{k_B T}{p - p_0}\right), \label{97} \end{equation}

\noindent where $p_0$ is the pressure at $\mathcal{P}_0$, and 2) a singular point $\mathcal{P}_0$ with weak interatomic interactions, and $R=0$, for which,

\begin{equation} R=-\kappa\left[\frac{k_B T}{p}-\left(\frac{k_B T}{p}\right)_0\right], \label{98} \end{equation}
 
\noindent where the quantity in parenthesis is evaluated at $\mathcal{P}_0$.
 
\par
Simplification results on using the scaled form:

\begin{equation} E=N \left(\frac{N}{V}\right)^a Y\left[\left(\frac{S}{V}\right)\left(\frac{N}{V}\right)^{-b}\right], \label{50} \end{equation}

\noindent where $a$ and $b$ are constant ''critical exponents'' and $Y()$ is a function of a single variable.  For the application in this paper, set $a=2/3$ and $b=1$, and Eq. (\ref{50}) becomes

\begin{equation} E=N\epsilon_F(\rho) Y(z), \label{60} \end{equation}

\noindent with Fermi energy {\cite{Pathria}}

\begin{equation} \epsilon_F(\rho) =\left(\frac{3^{2/3}\pi^{4/3}\,\hbar^2}{2m}\right)\rho^{2/3},  \label{64}\end{equation}

\noindent and $\rho=N/V$.  Physical constants have been included to set the energy scale; $\hbar$ is Planck's constant divided by $2\pi$, and $m$ is the atomic mass.  Eq. (\ref{60}) asserts that the internal energy per atom $E/N$, in units of the Fermi energy, is a function $Y(z)$ only of the entropy per atom in units of $k_B$, $z=S/N k_B$.  The ideal Fermi gas follows this form \cite{Pathria}, and this form is usually assumed even in the strongly interacting case \cite{Ho2004, Luo2009}.

\par
The values of $a$ and $b$ will depend on the spatial dimensionality, and the theory developed here should be applicable to strongly interacting systems of dimension other than three, with suitable adjustments of $a$ and $b$.

\section{GEOMETRIC EQUATION SOLUTION}

In this section, I present the solution to the geometric equation.  I develop the small $z$ (''superfluid'') part, the high $z$ (''normal'') part, and then connect these two parts at a second-order phase transition. 
 
\subsection{SMALL $z$ SOLUTION}

I start by solving the geometric equation in the regime of small $z$, where $Y(z)=Y_S(z)$, using a Puisseux series

\begin{equation} Y_S(z)=y_0 + y_1 \, z^{\alpha} + y_2 \, z^{2 \alpha} + \cdots, \label{100}\end{equation}

\noindent with exponent $\alpha$ and constant series coefficients $y_0$, $y_1$, $y_2$,... .  $(\alpha, y_0,  y_1)$ may be set freely, and the other series coefficients $y_i$ ($i\ge 2$) are uniquely determined by a series solution.  $\alpha>1$, $y_0>0$, and $y_1>0$ are necessary and sufficient conditions so that, for small positive $z$, $(E, T, p, g_{11}, g_{22}, g)$ are all positive.  With $\alpha>1$, $T\to 0$ as $S\to 0$, and Eq. (\ref{100}) is consistent with the third law of thermodynamics.  For $\alpha>1$ and $S\to 0$, we may also show that the heat capacity at constant volume goes to zero.

\par
$R$ follows directly from Eq. (\ref{30}):

\begin{equation} R=\frac{k_B\alpha V}{2S}\left[1-\left(\frac{(3 \alpha -5) y_1^2+10 y_0 y_2}{5 (\alpha -1) y_0 y_1}\right)x+O\left(x^2\right)\right], \label{110}\end{equation}

\noindent where

\begin{equation} x\equiv z^\alpha.\label{115}\end{equation}

\noindent Clearly as $z$ (and $S$) $\to0$, $R\to +\infty$.\footnote{ If $\alpha=2$, the leading term in the diverging $R$ is identical to that for the ideal Fermi gas.} The point with $z\to 0$ then corresponds to a singular point with $|R|\to\infty$, and the form of the geometric equation in Eq. (\ref{97}) is appropriate.  We have

\begin{equation}p_0=\left(\frac{\pi^{4/3}\,\hbar^2 y_0}{3^{1/3}m}\right)\rho^{5/3}, \label{117}\end{equation}

\noindent and

\begin{equation}-\kappa\left(\frac{k_B T}{p-p_0}\right)=\frac{-\kappa k_B \alpha V}{2S} \left[3+\left(\frac{3 y_2}{y_1}\right)x+O\left(x^2\right)\right]. \label{120}\end{equation}

\noindent Matching corresponding series terms in Eqs. (\ref{110}) and (\ref{120}), as required by the geometric equation Eq. (\ref{97}), now justifies the choice of the Puisseux series solution Eq. (\ref{100}), with the requirement

\begin{equation}\kappa=-1/3,\label{125}\end{equation}

\noindent and series coefficients $y_i$ ($i\ge 2$) determined uniquely in terms of $(\alpha, y_0, y_1)$.

\par
Eq. (\ref{97}) may be written as a third-order ordinary differential equation

\begin{equation}
\begin{array}{lr}
Y_S^{(3)}(z)= \\  \,\,\,
\displaystyle{\frac{2 Y_S'(z)^4+5 [y_0-2 Y_S(z)] Y_S'(z)^2 Y_S''(z) +10 [Y_S(z)-y_0] Y_S(z) Y_S''(z)^2}{5 [Y_S(z)-y_0] Y_S(z) Y_S'(z)}}.
\end{array}
\label{130}\end{equation}

\noindent Multiplying either $z$ or $Y_S(z)$ by constants leaves the form of Eq. (\ref{130}) unchanged, and two of the three required integration constants are thus scaling factors for $z$ and $Y_S(z)$.  These scaling factors are simply related to $y_0$ and $y_1$.  The third integration constant is the exponent $\alpha$.

\par
\begin{figure}[tpg]
\centering
\includegraphics[width=4.0in]{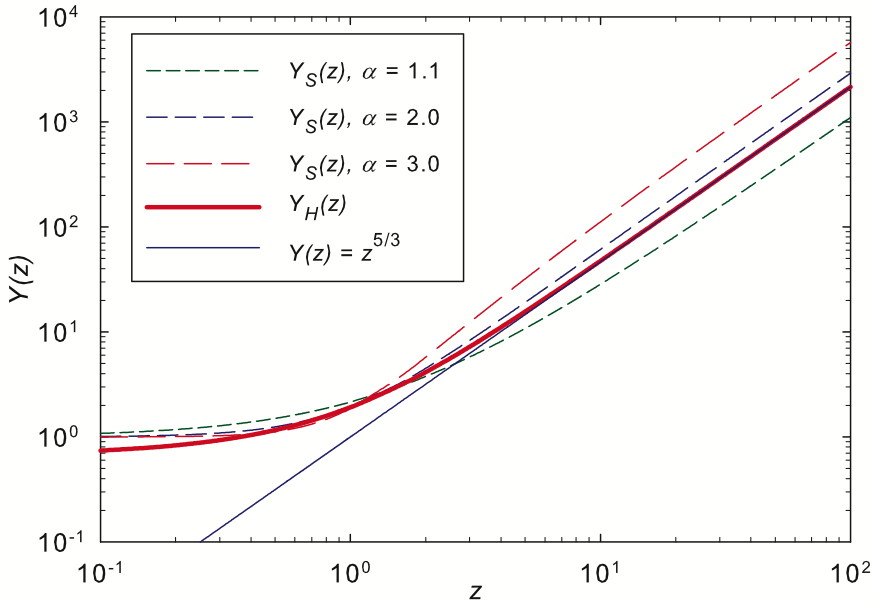}
\caption{$Y_S(z)$ versus $z$ for $\alpha=\{1.1, 2.0, 3.0\}$, and $Y_H(z)$ versus $z$.  On changing the free scaling factors these functions translate up and down and right and left, without changing shape, on a log-log scale.  The fine solid line shows $Y(z)=z^{5/3}$.  To get a phase transition, we must join a $Y_S(z)$ curve and a $Y_H(z)$ curve at the phase transition point $z=z_c$.}
\label{fig:1}
\end{figure}

To solve Eq. (\ref{130}) for $Y_S(z)$, we start by picking values for $(\alpha,y_0,y_1)$, and then generate an initial condition $\{z_0, Y_S(z_0), Y_S'(z_0), Y_S''(z_0)\}$ with the Puisseux series Eq. (\ref{100}).  For noninteger $\alpha$, this Puisseux series is not analytic at $z=0$, and so we must generate the initial condition about some small $z_0>0$.  I used $z_0=0.01$.  The numerical solution for $Y_S(z)$ indicates that $Y_S(z)$ is analytic for all $z\epsilon(0,\infty)$ for all the cases I tried representative of the interesting cases here.  Solutions are shown in Figure \ref{fig:1} for several values of $\alpha$.  Numerical solution shows that for large $z$, $Y_S(z)\propto z^{5/3}$ for all values of $\alpha$ I tried.  This is evident in Fig. \ref{fig:1}.

\par
I add that

\begin{equation} Y_S(z) = y_0 + y_1 \, z^{5/3}\label{140} \end{equation}

\noindent is an exact solution to the geometric equation for all values of $y_0$ and $y_1$.  This solution has

\begin{equation} R=\frac{5 k_B V}{6 S}, \label{145} \end{equation}

\noindent and is the toy example explored in Appendix 2.  The only drawback of this exact solution is that it does not fit the Duke experimental data very well.

\subsection{HIGH $z$ SOLUTION}

There is a second analytic solution $Y_H(z)$ to the geometric equation, and it corresponds to the singular point at $z\to\infty$, where $R\to 0$.  This second solution is, a priori, physically as logical as $Y_S(z)$, with fits to experimental data determining when and how to switch from one solution to the other.  There are two ways of joining $Y_S(z)$ and $Y_H(z)$: 1) The first way is via a first-order phase transition, with $Y_S(z)$ obtaining in the interval $z\epsilon\,[0,z_S)$, and $Y_H(z)$ obtaining in the interval $z\epsilon\,(z_H,\infty)$, with the constants $z_S$ and $z_H$ related by $z_{S}<z_{H}$.  The interval between $z_S$ and $z_H$ is physically excluded, with $z_H-z_S$ proportional to the latent heat per atom.  2) The second way to join the solutions is via a second-order phase transition, with joining point $z_S=z_H=z_c$.  I explore only this second method here.

\par
Assume that $Y_H(z)$ satisfies a Puisseux series valid for large $z$

\begin{equation} Y_H(z) =  \tilde{y} _ {-1} z^{\tilde{\alpha}} +  \tilde {y} _ 0 +  \tilde {y} _ 1  z^{-\tilde{\alpha}}  + \tilde {y} _ 2  z^{-2 \tilde{\alpha}}  +\cdots, \label{150} \end{equation}

\noindent where $\tilde{\alpha}$ is a constant exponent, $\tilde {y} _ {-1}$, $\tilde{y} _ {0}$, $\tilde{y} _ {1}$, $\tilde{y} _ {2}$, ... are constant series coefficients, with $\tilde{y}_{-1}$ and $\tilde{y}_0$ set freely, and $\tilde{y}_i$ ($i\ge 1$) determined by a series solution to the geometric equation.  In contrast to $\alpha$ in the $Y_S(z)$ solution, $\tilde{\alpha}$ may not be set freely.  Only $\tilde{\alpha}=5/3$ satisfies the geometric equation in the context of the Puisseux series.  For $\tilde{\alpha}=5/3$, and for large $z$, necessary and sufficient conditions for positive $(E, T, p, g_{11}, g_{22}, g)$ are $\tilde{y}_{-1}>0$ and $\tilde{y}_0>0$.

\par
With $\tilde{\alpha}=5/3$, the series for $Y_H(z)$ yields

\begin{equation} R = \frac{k_B V}{S} \left[\frac{5}{6}  + \frac{35\tilde{y}_{1}}{6\tilde{y}_0}\tilde{x} + O(\tilde{x}^2) \right], \label{180} \end{equation}

\noindent and

\begin{equation} \frac{k_B T}{p} =\frac{k_B V}{S}\left[\frac{5}{2}-\frac{5\tilde{y}_0}{2 \tilde{y}_{-1}}\tilde{x}+O(\tilde{x}^2)\right] \label{190}, \end{equation}

\noindent where

\begin{equation} \tilde{x}\equiv z^{-\alpha} \label{194}. \end{equation}

\noindent Clearly as $z$ (and $S$) $\to\infty$, $R\to 0$, and we have a singular point of the second type above, with a geometric equation of the form in Eq. (\ref{98}).  At the singular point

\begin{equation} \left(\frac{k_B T}{p}\right)_0 = 0, \label{196} \end{equation}

\noindent and the geometric equation is

\begin{equation} R=-\kappa\left(\frac{k_B T}{p}\right).\label{198} \end{equation}

\noindent The series solution to Eq. (\ref{198}) yields $\kappa=-1/3$, the same universal value as in the previous section.  The series solution also yields the coefficients $\tilde{y_i}$ $(i\ge1)$ in terms of $\tilde{y}_{-1}$ and $\tilde{y}_0$.

\par
The full third-order ordinary differential equation is

\begin{equation} Y_H^{(3)}(z)=\displaystyle{\frac{2 Y_H'(z)^4-10 Y_H(z) Y_H'(z)^2 Y_H''(z) +10 Y_H(z)^2 Y_H''(z)^2}{5 Y_H(z)^2 Y_H'(z)}}. \label{200}\end{equation}

\noindent This equation is invariant under multiplication of $z$ or $Y_H(z)$ by scaling factors.  We may solve it numerically by generating an initial condition with a series for given $(\tilde{\alpha}=5/3, \tilde{y}_{-1}, \tilde{y}_0)$ and some given small $\tilde{x}$.  A full solution is shown in Fig. \ref{fig:1}, where changing $(\tilde{y}_{-1}, \tilde{y}_0)$ results only in vertical and horizontal translations on the log-log scale.

\par $\tilde{\alpha}\ne5/3$ leads to incompatible series for the geometric equation.  Furthermore, as I show in Appendix 2, the only exponent which allows one to trap a gas with no leakage at the edge is $\tilde{\alpha}=5/3$.  I explore no other values for $\tilde{\alpha}$ here.

\subsection{JOIN AT THE PHASE TRANSITION}

The functions $Y_S(z)$ and $Y_H(z)$ must be joined to get the complete solution.  This joining results inevitably in a phase transition.  I consider only the possibility of a second-order phase transition, at a single value $z=z_c$.  However, the joining method could readily be extended to first-order phase transitions.

\par
Define the quantities per volume $e(s, \rho)=E(S,N,V)/V$, $s=S/V$, and $\rho=N/V$.  By Eq. (\ref{60}),

\begin{equation} e(s,\rho)=\rho\,\epsilon_F(\rho) Y(z). \label{210}\end{equation}

\noindent Since $\{T, \mu, p\}=\{e_{,s},e,_{\rho},2e/3\}$, and since $z$ can be written as $z = s/\rho k_B$, we have

\begin{equation} k_B T = \epsilon_F(\rho)Y'(z), \label{220}\end{equation}

\begin{equation} p = \frac{2}{3} \rho\,\epsilon_F(\rho)Y(z), \label{230}\end{equation}

\noindent and

\begin{equation}\mu= \frac{5 p}{2 \rho} - k_B T z. \label{240}\end{equation}
 
\par
Consider now joining $Y_S(z)$ and $Y_H(z)$ at some point with $z=z_c$ common to both curves, corresponding to the absence of molar latent heat.  For either a first or a second-order phase transition, we require continuous $\{T,\mu, p\}$.  Adding the condition of continuous $z$, Eq. (\ref{240}) now requires also continuous density $\rho$.  The joining conditions are thus

\begin{equation} Y_S(z_c)=Y_H(z_c), \label{250} \end{equation}

\noindent and

\begin{equation} Y'_S(z_c) = Y'_H(z_c). \label{260} \end{equation}

\noindent These two conditions allow us to write $(\tilde{y}_{-1}, \tilde{y}_{0})$ uniquely in terms of $(\alpha,y_0,y_1,z_c)$.

\par
We may now generate a complete solution for $Y(z)$.  I tested several solutions, and found that in each case $(E, T, p, g_{11}, g_{22}, g)$ are all positive over the full range of $z$ from zero to very large, as required by thermodynamics.

\section{FITS TO DUKE DATA}

In this section, I fit the Local Density Approximation (LDA) in this paper to the Duke experiment \cite{Luo2007,Luo2009}.  The Duke experiment gathered data, including error bars, for the total energy, total entropy, and total number of atoms $(E_t, S_t, N_t)$, respectively.  The experimental system consisted of $N_t\sim 1.3\times 10^5$ fermionic $^6$Li atoms in a 50:50 mixture of the two lowest energy hyperfine states.  This mixture was confined in a laser trap with a Gaussian potential, in a magnetic field tuned just above a broad Feshbach resonance.

\par
To compare an LDA to experiment requires trap integration over the potential energy per atom $U(\vec{r})$.  Details are described in Appendix 2.  For large $z$, $Y_H(z)$ has a power law limiting form $\propto z^{5/3}$, by Eq. (\ref{150}).  This form yields a nice trap boundary ($\rho=0$ and $z\to\infty$) in every direction.

\par
A complication with the power law limiting form $\propto z^{5/3}$ is that it adds a boundary pressure integral to the usual virial theorem \cite{Thomas2005}.  It is assumed in most applications that the gas at the edge of the trap becomes an ideal gas, obeying the Sackur-Tetrode equation for which the pressure approaches zero as the density goes to zero at constant temperature.  This assumption appears to work very well in experiments, but there is no guarantee that unitary thermodynamics actually behaves like an idea gas in the limit $z\rightarrow\infty$.  I examine this issue in some detail in Appendix 3, but this point does not affect the analysis given below.

\par
The Duke group determined the trap entropy $S_t$ in three different ways.  In conjunction with their trap energy $E_t$ data ''$E_{840}/E_F$'' and their $N_t$ values, I analyzed their $S_t$ data ''$S_{1200}^{**}/k_B$,'' since it has the smallest error bars.  Details of calculating $\chi^2$ from $(\alpha,y_0,y_1,z_c)$ are given in Appendix 2.  Proceed by picking fixed values of $(\alpha,z_c)$, and adjust the scaling constants $(y_0,y_1)$ for $Y_S(z)$ to minimize $\chi^2$.  The scaling constants $(\tilde{y}_{-1},\tilde{y}_{0})$ for $Y_H(z)$ follow from $(\alpha, y_0, y_1,z_c)$ by Eqs. (\ref{250}) and (\ref{260}), and there is no need to vary them separately.

\begin{figure}[tpg]
\centering
\includegraphics[width=4.0in]{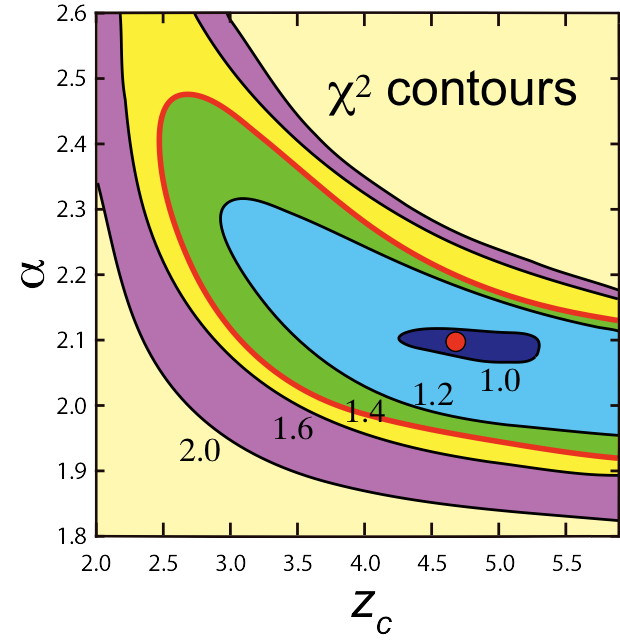}
\caption{$\chi^2$ contours as a function of $(\alpha,z_c)$.  The very best fit, with $\chi^2=0.95$, is denoted by a red dot.  The bold red curve indicates $\chi^2=1.4$, the approximate limit of admissible fits.  In conjunction with their $E_t$ and $N_t$ values, I analyzed the Duke group $S_t$ data ''$S_{1200}^{**}/k_B$,'' since it had the smallest error bars.}
\label{fig:2}
\end{figure}

\par
\begin{figure}[tpg]
\centering
\includegraphics[width=4.0in]{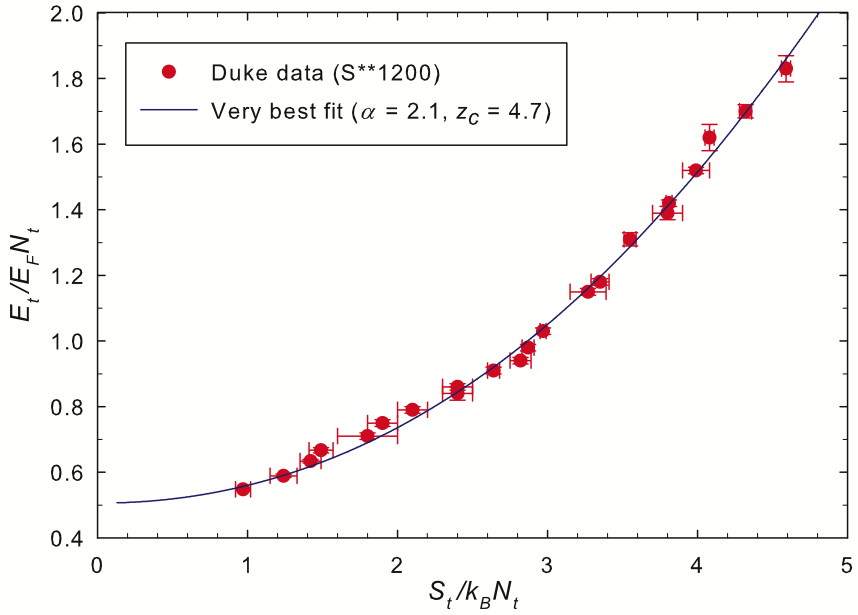}
\caption{The very best fit to the Duke data set, with $\alpha=2.1$, $z_c=4.7$, and $\chi^2=0.95$.}
\label{fig:3}
\end{figure}

Figure \ref{fig:2} shows this minimum $\chi^2$ as a function of $(\alpha,z_c)$.  With increasing $z_c$, the contours become horizontal as $Y_S(z)$ encompasses an increasing fraction of the experimental data.  Working with larger values of $z_c$ is not very revealing, and a reliable upper limit on $z_c$ was difficult to determine.  A lower limit for $z_c$ was more easily located.  Fig. \ref{fig:2} shows marginal fits having $\chi^2\sim 1.4$, with $z_c$ as small as $2.5$.  Values of $z_c$ smaller than 2.5 would appear to be clearly inconsistent with my analysis of this data set.  Figure \ref{fig:3} shows the very best fit with $\alpha=2.1$, $z_c=4.7$, $y_0=0.277$, and $y_1=0.0735$, with $\chi^2=0.95$.  

\par
Essential in the discussion of unitary thermodynamics is the Bertsch parameter, defined by \cite{Haussmann2008}

\begin{equation} \xi_B\equiv\frac{5 E}{3N\epsilon_F(\rho)}, \label{270}\end{equation}

\noindent evaluated in the limit $T\rightarrow 0$.  $\xi_B$ is expected to have a universal value, the same for all unitary thermodynamic gases.  My LDA yields a value for $\xi_B$; from Eq. (\ref{100}),

\begin{equation} \xi_B=\frac{5}{3}y_0. \label{280}\end{equation}

\begin{figure}[tpg]
\centering
\includegraphics[width=4.0in]{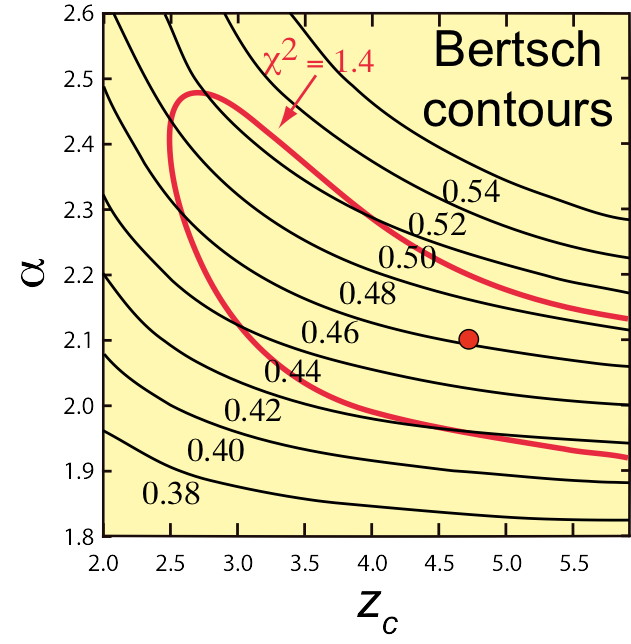}
\caption{The Bertsch parameter $\xi_B$ as a function of $(\alpha,z_c)$.  The red dot corresponds to the very best fit, and the bold red curve curve to $\chi^2=1.4$.  }
\label{fig:4}
\end{figure}

\noindent Figure \ref{fig:4} shows the results.  The very best fit shown in Figure \ref{fig:3} corresponds to $\xi_B=0.462(40)$, with error bar estimated from Figure \ref{fig:4}.  This $\xi_B$ overlaps with the value $\xi_B = 0.435(15)$ determined in the Duke experiment with speed of sound measurements \cite{Luo2009}.  The bold red curve curve in Fig. \ref{fig:4} corresponds to $\chi^2=1.4$, and the Duke value for $\xi_B$ fits comfortably in this zone.  For example, $\alpha=2.1$, $z_c=3.0$, had $\chi^2=1.44$, and $\xi_B=0.434$, a value for $\xi_B$ in agreement with the Duke experiment.  However, these values are higher than the value $\xi_B=0.376(4)$ reported by Ku {\it et al.} \cite{Ku2012}.

\par
\begin{figure}[tpg]
\centering
\includegraphics[width=4.0in]{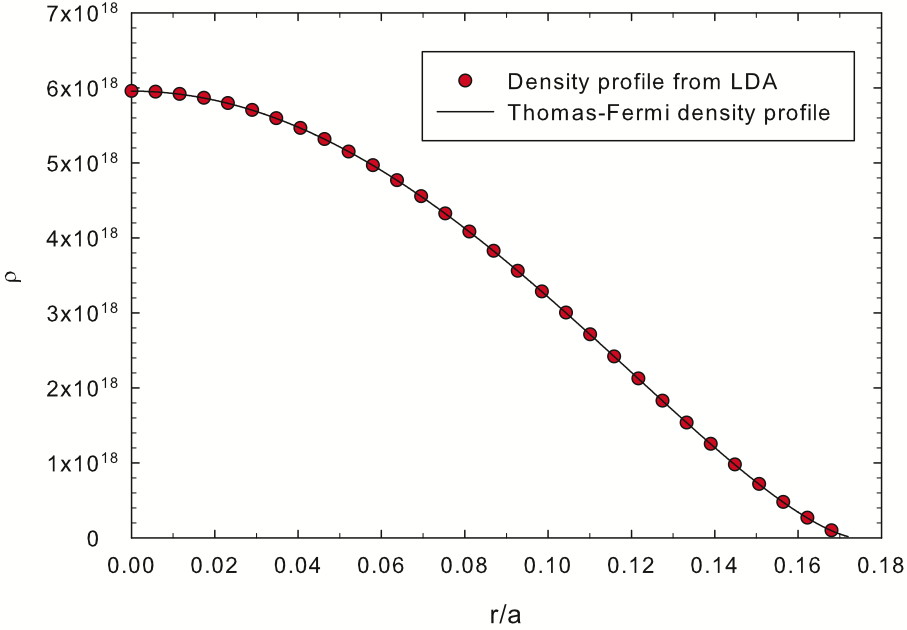}
\caption{The predicted trap density profile at effectively zero temperature calculated with my very best fit LDA.  Also shown is the corresponding Thomas-Fermi density profile.}
\label{fig:5}
\end{figure}

Also of considerable interest in trapped Fermi systems are density profiles $\rho(\vec{r})$.  I calculated a theoretical density profile for the Duke Gaussian trap for $T=0.01\mu$K (much lower than any temperature encountered in the experiment), $N_t=1.3\times 10^5$, and my very best fit equation of state in Fig. \ref{fig:3}.  These parameters, and the Gaussian potential in Eq. (\ref{2050}), require $\mu_0/U_0=0.0586$.  The corresponding density profile is shown in Figure \ref{fig:5} as a function of the scaled distance $\tilde{r}$ along a radial path from the center of the trap (defined in Appendix 2).  With appropriate scaling for the distances, the trap is spherically symmetric.  To calculate trap properties, I evaluated the LDA at about $30$ equally spaced $\tilde{r}$ values, interpolated a curve through them, and integrated.   Also shown in Fig. \ref{fig:5} is the Thomas-Fermi density profile in Eq. (\ref{2040}).  Remarkably, the Thomas-Fermi density profile fits that from my LDA almost exactly!

\par
I also tried fits using only one segment $Y_S(z)$ ($z_c\rightarrow\infty$).  But such fits did not produce values of $\chi^2$ as low as those with the two-segment method.  Therefore, although it proved difficult to determine an upper limit on $z_c$ with the two-segment method with this data set, there clearly is one, indicating the existence of a phase transition.

\par
\begin{figure}[tpg]
\centering
\includegraphics[width=4.0in]{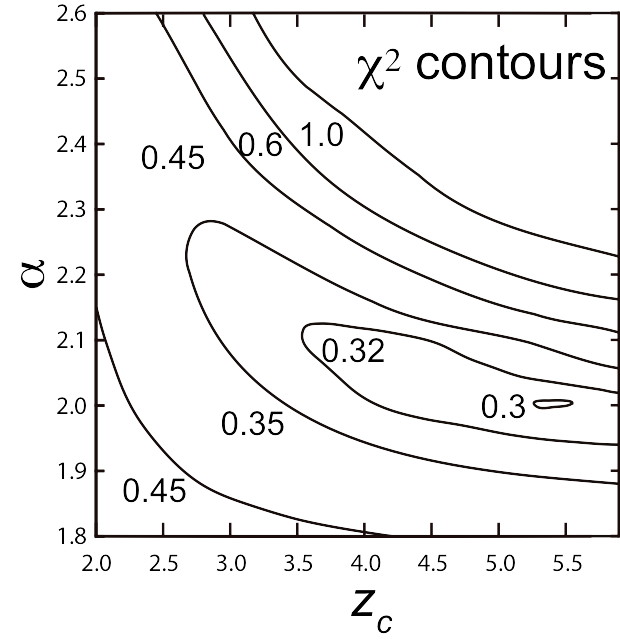}
\caption{Analysis using the entropy data ''$S_{1200}^{*}/k_B$'' in \cite{Luo2009}.  The figure shows $\chi^2$ contours, which are rather undiscriminating with respect to variations in the fitting parameters.  A broad range of $(\alpha,z_c)$ values has $\chi^2$ well under unity.  Because of its broad error bars, the analysis with this entropy was disfavored over that with ''$S_{1200}^{**}/k_B$."}
\label{fig:6}
\end{figure}

Cao {\it et al.} \cite{Cao2011} found that to make a successful temperature calibration with the Duke data, it was necessary to use the entropy data labeled ''$S_{1200}^{*}/k_B$,'' \cite{Luo2009} which was corrected for the finite interaction strength in the weakly interacting gas.  The data set  ''$S_{1200}^{**}/k_B$'' featured in my paper does not have this correction.  However, ''$S_{1200}^{*}/k_B$'' has error bars over twice as large on the average as ''$S_{1200}^{**}/k_B$,'' and this leads to problems with the $\chi^2$ analysis.  Figure \ref{fig:6} shows $\chi^2$ as a function of $(\alpha,z_c)$ calculated using the data set  ''$S_{1200}^{*}/k_B$.''  As can be seen, the result is rather indiscriminating with regard to different values of the fitting parameters, with a large range of values producing $\chi^2$ well less than unity.  Such values indicate either that I have used too many fitting parameters for the data points (not the case in Fig. \ref{fig:2}), or that the error bars in \cite{Luo2009} were perhaps too conservatively reported.  For this reason, I featured ''$S_{1200}^{**}/k_B$'' in my analysis.

\section{DISCUSSION}

A feature of my presentation is the power law type analysis, determining critical exponents and scaling factors by fits to experimental data.  This contrasts with techniques in which the equations of state are calculated from the atoms up with quantum Monte Carlo simulation, for example \cite{Giorgini2008}.

\par
Luo and Thomas \cite{Luo2009} performed a power law analysis on their data, with interesting results.  They used a function in two sections, separated by a phase transition point $z=z_c$, which served as one of the fitting parameters.  However, these fitting functions were ad hoc functions for the thermodynamics measured in their trap.  There was no explicit connection to any LDA, nor any trap integration to achieve the fit.  This makes their results difficult to compare with the try in this paper.

\section{CONCLUSIONS}

\par
In conclusion, I have presented a solution for unitary thermodynamics and fit it to experimental data taken in a Gaussian trap.  My equation uses a new technique based on the metric geometry of thermodynamics.  This thermodynamic approach takes implicit advantage of large correlation lengths, the very element which makes traditional calculations based on statistical mechanics difficult to do for strongly interacting systems.  The approach in this paper is general in its ideas and applications, with a direct application method, and fits for a critical exponent and scaling constants in the style of critical phenomena.  The resulting theory is a scaled fundamental equation for the internal energy per atom in units of the Fermi energy $Y(z)$, where $z$ is the entropy per atom in units of $k_B$.  The theoretical solution comes in two sections, a "superfluid" phase $Y_S(z)$ from $0<z<z_c$ and a "normal" phase $Y_H(z)$ from $z_c<z<\infty$, joined at a second-order phase transition at $ z_c$.  Fits to the data of the Duke experiment were excellent, with $\chi^2\sim 1$.  The fits were rather insensitive to the position of $z_c$.  However, two segment fits worked better than one segment fits using only $Y_S(z)$, so I conclude that I have indicated the existence of a phase transition, though have had a difficulty actually locating where it is.

\par
I thank John Thomas and Wilhelm Zwerger for useful communications, and Horst Meyer for encouragement.  I also thank George Skestos for travel support.

\section{APPENDIX 1: GEOMETRIC EQUATION}

In this Appendix, I justify the uniqueness of the geometric equation

\begin{equation}R=-\frac{\kappa}{\phi}, \label{1000}\end{equation}

\noindent with $\kappa$ a dimensionless constant of order unity, and the thermodynamic potential per volume, in units of $k_B$,

\begin{equation}\phi=\frac{p}{k_B T}. \label{1010}\end{equation}

\noindent $\phi$ is always employed in conjunction with an appropriate background subtraction, depending on the singular point around which the solution is generated.  There are two types of singular points: 1) singular points with $|R|\to\infty$, where the background subtraction is made from $\phi$, and 2) singular points with $R\to 0$, where the background subtraction is made from $1/\phi$.

\par
Singular points with $|R|\to\infty$ are encountered at critical points, where intermolecular interactions strongly organize the atoms.  Singular points with $R\to 0$ are encountered in near ideal gases where intermolecular interactions have little effect.   Considerations of simplicity, units, mathematical viability, and universality of $\kappa$ turn out to give us little choice about the correct form of $\phi$ and in the subtraction of the nonsingular part.  Although I state my arguments in the context of the specific physical problem here, such arguments have been used in quite different physical settings: the simple critical point \cite{Ruppeiner1991}, galaxy clustering \cite{Ruppeiner1996}, corrections to scaling \cite{Ruppeiner1998}, the paramagnetic ideal gas \cite{Kaviani1999}, and gases with power law interactions \cite{Ruppeiner2005}.  In all these cases, the geometric equation takes the same form as that presented here, and with the same prescription for the background subtraction.

\par
Consider a singular point with $|R|\to\infty$, and start the discussion with hyperscaling.  Widom \cite{Widom74} argued that, near a critical point, average fluctuations in the free energy $\Phi$ in a volume of size $\xi^3$ should be $\sim k_B T$.  If we further take these fluctuations to be the singular part of the free energy itself, we get the hyperscaling assumption:

\begin{equation}\frac{\Phi-\Phi_0}{V}\sim\frac{k_B T}{\xi^3},\label{1015}  \end{equation}

\noindent where $\Phi_0$ denotes the value of $\Phi$ at the critical point, which we must subtract to get the singular properties corresponding to $\xi\to\infty$.

\par
To connect $\xi$ to thermodynamic properties in hyperscaling, it is necessary to express $\Phi$ in thermodynamic terms.  Widom \cite{Widom74} was not explicit in this regard, but he nevertheless used Eq. (\ref{1015}) to write the hyperscaling exponent scaling relation $2-\alpha=d\,\nu$ between the heat capacity (at constant volume) exponent $\alpha$ and the correlation length exponent $\nu$, in spatial dimension $d$.  Goodstein \cite{Goodstein} made an argument similar to Widom's in spirit, and picked $\Phi$ as the Gibbs free energy.  He stated, however, that ''it will turn out not to make any difference which energy function we choose.''

\par
Such looseness with the precise definition of $\Phi$ is sufficient if we are interested only in the hyperscaling exponent relation, but in the context of the geometric equation we must be more precise.  Bringing in the thermodynamic curvature $R$ forces a sharpening of the argument.  In the critical regime, $R$ connects to $\xi$ via $|R|\sim\xi^3$, a proportionality resulting both from direct calculations in a number of cases (see \cite{Ruppeiner1995, Johnston2003, Ruppeiner2010} for review), and from a covariant theory of thermodynamic fluctuations \cite{Ruppeiner1983a, Ruppeiner1983b, Diosi1985}.

\par
Replacing $\xi^3$ with $R$ in Eq. (\ref{1015}) leads to

\begin{equation} R=-\kappa\left(\frac{k_B T V}{\Phi-\Phi_0}\right), \label{1020}\end{equation}

\noindent with $\kappa$ a dimensionless constant of order unity and ''$\sim$'' replaced by ''$=$''.  $R$ has units of volume per molecule, so $\Phi$ must have units of energy.  I will try all possible $\Phi$'s constructed from the four free energy building blocks $\{E, -T S, p V, -\mu N\}$, each extensive and each with units of energy.  I find that of the 14 nontrivial possibilities, only $\Phi=pV$, and cases with $\Phi\propto pV$, are viable.

\par
The specific procedure for testing $\Phi$'s is: 1) Pick a singular point where, on physical grounds, we expect $|R|\to\infty$, and guess some physically motivated series yielding the thermodynamics.  Such a series should contain undetermined coefficients, to be evaluated by series solution of the geometric equation.  2) Expand $R$ in terms of this series.  3) Construct a candidate $\Phi$ from the four building blocks $\{E, -T S, p V, -\mu N\}$, and evaluate it at the singular point to get $\Phi_0$.  4) Construct the series for $-\kappa k_B T V/(\Phi-\Phi_0)$, with $\kappa$ undetermined.  5) Equate this series to the one for $R$ and see whether the resulting $\kappa$ is universal.  By ''universal,'' I mean independent of the specific constants found in the solution.  6) Repeat this procedure until all candidate $\Phi$'s have been tried.

\par
For unitary thermodynamics, we expect $R\to +\infty$ at $z=0$, characteristic of the noninteracting Fermi gas \cite{Mrugala1990, Oshima1999}.  Try a Puisseux series

\begin{equation} Y(z)=y_0+y_1 \, z^{\alpha }+y_2 \, z^{2 \alpha}+\cdots \label{1030}\end{equation}

\noindent to calculate the energy,

\begin{equation} E=N\epsilon_F(\rho) Y(z). \label{1035} \end{equation}

\noindent Eqs. (\ref{30}), (\ref{1030}), and (\ref{1035}) lead to

\begin{equation} R=\frac{k_B\alpha V}{2S}\left[1-\left(\frac{(3 \alpha -5) y_1^2+10 y_0 y_2}{5 (\alpha -1) y_0 y_1}\right)x+O\left(x^2\right)\right], \label{1040}\end{equation}

\noindent where $x\equiv z^\alpha$.

\par
As the first free energy candidate, try $\Phi = p V$, the case featured in this paper.  Since $\{T,p\}=\{E_{,S}\,,-E_{,V}\}$, Eqs. (\ref{1030}) and (\ref{1035}) lead to

\begin{equation} p_0 = \frac{2}{3} c y_0 \rho^{5/3}, \label{1045} \end{equation}

\noindent with

\begin{equation} c = \frac{3^{2/3}\pi^{4/3}\,\hbar^2}{2m}, \label{1047} \end{equation}

\noindent and

\begin{equation}\phi_{sing}\equiv\frac{p-p_{0}}{k_B T} = \frac{2S}{k_B \alpha V}  \left[\frac{1}{3}-\left(\frac{y_2}{3 y_1}\right)x+O\left(x^2\right)\right].  \label{1050}\end{equation}

\noindent Thus

\begin{equation}-\kappa=R\, \phi_{sing}=\frac{1}{3}-\left[\frac{(3 \alpha -5) y_1^2+5 (\alpha +1) y_0 y_2}{15 (\alpha -1) y_0 y_1)}\right]x + O\left(x^2\right).  \label{1053}\end{equation}

\noindent Clearly, $\kappa$ must take the universal value $\kappa=-1/3$, regardless  the values of $\alpha$ and the series coefficients.  Setting the first-order term on the right-hand side of Eq. (\ref{1053}) to zero requires $y_2=-(3 \alpha -5) y_1^2/[5 (\alpha +1) y_0]$, and setting higher-order terms to zero uniquely determines all the series coefficients in terms of $\{\alpha,y_0,y_1\}$.

\par
All 14 possible nonzero ways of creating a free energy $\Phi$ are presented in Table 1.  The form of the fundamental equation in Eq. (\ref{1035}) leads to $E=3pV/2$.  Generally, we also have the Gibbs-Duhem equation $E = T S - p V + \mu N$.  Hence, a number of the candidate $\Phi$'s in Table 1 have $\Phi\propto pV$, and they all lead to the same form of the geometric equation as $\Phi = p V$.  Cases with $\Phi\propto p V$ are the only ones with universal $\kappa$.  All other cases have $\kappa$ depending on $\alpha$, and are hence unacceptable.

\begin{table}
\begin{tabular}{|c|c|c|}
\hline
\hline
$\Phi$ & $\Phi_{0}$ & $R\,\phi_{sing}$\\
\hline
$E=\frac{3}{2}pV$ & $c\,y_0\,\rho^{5/3}V$ & $1/2+O(x)$ \\
$-T S$ & $0$ & $-\alpha/2 +O(x)$ \\
$p V$ & $2c\,y_0\,\rho^{5/3}V/3$ & $1/3 + O(x) $ \\
$-\mu N$ & $-5c\,y_0\,\rho^{5/3}V/3$ & $(3\alpha-5)/6 + O(x)$ \\
$E - T S$& $c\,y_0\,\rho^{5/3}V$ & $(-\alpha + 1)/2+O(x)$ \\
$E + p V=\frac{5}{2}pV$ & $5c\,y_0\,\rho^{5/3}V/3$& $5/6 + O(x)$ \\
$E - \mu N$ & $-2c\,y_0\,\rho^{5/3}V/3$ & $(3\alpha-2)/6+O(x)$ \\
$- T S + p V$ & $2c\,y_0\,\rho^{5/3}V/3$ & $(-3\alpha + 2)/6+O(x)$ \\
$- T S - \mu N = -\frac{5}{2}pV$ & $-5c\,y_0\,\rho^{5/3}V/3$ & $-5/6+O(x)$ \\
$p V - \mu N$ & $-c\,y_0\,\rho^{5/3}V$ & $(\alpha-1)/2 +O(x)$  \\
$E - T S + p V$& $5c\,y_0\,\rho^{5/3}V/3$ & $(-3\alpha+5)/6+O(x)$ \\
$E - T S - \mu N = -pV$ & $-2c\,y_0\,\rho^{5/3}V/3$ & $-1/3+O(x)$ \\
$E + p V - \mu N$ & $0$ & $\alpha/2+O(x)$ \\
$ - T S + p V - \mu N = - \frac{3}{2}pV$ & $-c\,y_0\,\rho^{5/3}V$ & $-1/2+O(x)$ \\
\hline
\end{tabular}
\centering
\caption{\label{} All $14$ nontrivial free energies $\Phi$ constructed from $\{E, -T S, p V, -\mu N\}$.  Also shown are $\Phi_0$ and $R\,\phi_{sing}$ for small $z$, where $\phi_{sing}\equiv (\Phi-\Phi_0)/k_B T V$.  A number of cases have $\Phi\propto pV$, of which all have values of $\kappa$ independent of solution.  The other cases have $\kappa$ depend on $\alpha$, and are hence nonuniversal.}
\end{table}

\par
Finally, turn attention briefly to singular points with $R\to 0$.  For unitary thermodynamics, such singular points correspond to $z\to\infty$.  For singular points with $R\to 0$, we must subtract $1/\phi_0$ from $1/\phi$ to get $R=0$, where $\phi_0$ is $\Phi$ evaluated at the singular point.  I will not present an explicit analysis of this case, since it is clear that we must have $\Phi = pV$ to be consistent with the case above with $z\to 0$.  Likewise, it turns out that there is little choice about the appropriate background subtraction.  The geometric equation for singular points with $R\to 0$ is thus

\begin{equation} R = -\kappa\left[\frac{k_B T}{p}-\left(\frac{k_B T}{p}\right)_0\right], \label{1060}\end{equation}

\noindent were the quantity in parentheses on the right-hand side is evaluated at the singular point.  As shown in Section 3.2, this equation is entirely solvable, with $\kappa=-1/3$, the same as for the small $z$ solution.

\section{APPENDIX 2: LDA $\to$ TRAP}

\par
In this Appendix, I present the basics of connecting the uniform thermodynamics (LDA) to the measured overall properties of a fluid in a trap.  This topic was discussed by Haussmann and Zwerger \cite{Haussmann2008}, and I add to their discussion in this Appendix a simple ad hoc power law toy example, which I fit to experimental data from the Duke University group \cite{Luo2009}.  This fit is not expected to be particularly good, but it does raise some useful points for discussion.

\par
Consider a thermodynamic system in a trap where an atom at position $\vec{r}$ experiences a known external potential energy per atom $U(\vec{r})$ in addition to the net potential energy contributed by the other atoms in the system.  Let $U(\vec{r})$ have a minimum $U(0)=0$, and increase monotonically with $r=|\vec{r}|$ in all directions.  Assume that the LDA is also known and is given by the fundamental equation $E(S,N,V) = V e(s, \rho)$, where $e(s,\rho)=E/V$ is the internal energy per volume, $s=S/V$, and $\rho=N/V$.  The temperature and the thermodynamic chemical potential are given by $\{T,\mu\}=\{e,_s, e,_{\rho}\}$.  Logically, in going from an LDA to a trap thermodynamics, the volume variable $V$ gets replaced by the potential energy per atom $U(\vec{r})$ \cite{Hansen}.

\par
$T$ and the total chemical potential $\mu_0 = \mu + U(\vec{r})$ are both constant throughout the trap.  The basic question is: for a known LDA and given $T$, $\mu_0$, and $U(\vec{r})$, what are the energy $E_t$, entropy $S_t$, and number of atoms $N_t$ in the trap?  To determine this transformation $\{T, \mu_0, U(\vec{r})\}\to\{E_t, S_t, N_t\}$, proceed as follows: 1) Determine the local $\mu=\mu_0 - U(\vec{r})$ for all $\vec{r}$.  2) Algebraically solve $(T=e,_s, \mu=e,_{\rho})$ for $(s(\vec{r}), \rho(\vec{r}))$ in terms of $(T, \mu_0)$ for all $\vec{r}$.  3) Identify the boundary (or edge) of the trap by finding the surface over which $\rho=0$.  4) Integrate over the volume of the trap out to the edge:

\begin{equation} E_t=\int\left[e(s(\vec{r}),\rho(\vec{r})) + \rho(\vec{r})U(\vec{r})\right]\,d^3r,\label{1970}\end{equation}

\begin{equation} S_t=\int s(\vec{r})\,d^3r,\label{1980}\end{equation}

\noindent and

\begin{equation} N_t=\int \rho(\vec{r})\,d^3r.\label{1990}\end{equation}

\par
Several questions come up at the edge of the trap.  In some direction, is there a {\it finite} distance $r$ where $\rho\to 0$, or does $\rho$ instead slowly peter out only as $r\to\infty$?  Need $U(\vec{r})$ diverge to infinity to contain the atoms in the trap?  Does $\rho\to 0$ imply $z\to\infty$?  Do the integrals above for $(E_t,S_t,N_t)$ converge?

\par
To illuminate these issues, consider a simple toy example based on the scaled equation of state in Eq. (\ref{60}):

\begin{equation} e(s, \rho) = \rho\,\epsilon_F(\rho) Y(z), \label{2000}\end{equation}

\noindent with $z=s/k_B\rho$.  Take a power law,

\begin{equation} Y(z) = y_0 + y_1 z^{\alpha},\label{2005}\end{equation}

\noindent with constants $y_0>0$, $y_1>0$, and $\alpha>1$.  $T=e,_s$ yields

\begin{equation}  \rho(T,z) = \frac{2\sqrt{2}}{3 \pi^2} \left(\frac{m k_B T}{\alpha y_1z^{\alpha-1}\,\hbar^2}\right)^{3/2}. \label{2010}\end{equation}

\noindent The trap edge $\rho\to 0$ clearly corresponds to $z\to\infty$, physically reasonable since the space available to an atom, and hence the entropy per atom, grows without limit as $\rho\to 0$.  The condition $e_{,\rho}=\mu = \mu_0 - U(\vec{r})$ and Eq. (\ref{2010}) yields

\begin{equation} -k_B T z + \frac{5 k_B T z}{3 \alpha} + \frac{5 y_0 k_B T }{3\alpha y_1 z^{\alpha-1}} = \mu_0 - U(\vec{r}).\label{2020}\end{equation}

\par
If $\alpha=5/3$, the terms linear in $z$ in Eq. (\ref{2020}) cancel, and Eqs. (\ref{2010}) and (\ref{2020}) leads to

\begin{equation} s(T,\mu_0,\vec{r}) = \frac{2}{5}\sqrt{\frac{6}{5}}\frac{k_B}{\pi^2}\left[\frac{m k_B T}{\hbar^2 y_1}\right]^{3/2},\label{2030}\end{equation}

\noindent with density profile

\begin{equation} \rho(T,\mu_0,\vec{r}) = \frac{2}{5}\sqrt{\frac{6}{5}}\frac{1}{\pi^2}\left\{\frac{m[\mu_0-U(\vec{r})]}{\hbar^2 y_0}\right\}^{3/2}.\label{2040}\end{equation}

\noindent This $\rho(T,\mu_0,\vec{r})$ is independent of $T$ and follows the Thomas-Fermi density profile \cite{Giorgini2008}.  $e(T,\mu_0,\vec{r})$ now follows from $e(s,\rho)$ since $s$ and $\rho$ are known at this point in terms of $(T,\mu_0,\vec{r})$.  Clearly, a real valued $\rho$ at $\vec{r}$ requires $\mu_0>U(\vec{r})$, which is consistent with Eq. (\ref{2005}).  As $r$ increases from zero in some direction, $U(\vec{r})$ increases until $U(\vec{r}) = \mu_0$, assuming that $\mu_0$ is not too big.  When $U(\vec{r})\to \mu_0$, we get $\rho\to 0$ and $z\to\infty$, corresponding to the trap edge.  Assuming that we have such a trap edge in every direction, the integrals for $E_t$, $S_t$, and $N_t$ will all converge, since the volume of integration is finite.

\par
With $\alpha>1$, but $\alpha\ne 5/3$, the linear $z$ terms dominate in Eq. (\ref{2020}) as $z\to\infty$.  But diverging $z$ now requires $|\mu_0-U(\vec{r})|\to\infty$, which clearly cannot happen since $0\le U(\vec{r})\le\mu_0$, and $\mu_0$ has been set to some fixed value characteristic of the entire system.  $\alpha=5/3$ is thus the only exponent leading to a clear trap edge, and with no need for any infinity in $U(\vec{r})$ to confine the atoms.

\par
As a practical exercise, let me compare the toy LDA in Eq. (\ref{2005}), with $\alpha=5/3$, with the experimental trap data collected by the Duke group \cite{Luo2009}.  The Duke group used an optical trap with a Gaussian potential

\begin{equation} U(\vec{r})=U_0\left[1-\mbox{exp}\left(-2\,\tilde{r}^2\right)\right], \label{2050}\end{equation}

\noindent where $U_0=10\,\mu$K$\,k_B$, $\tilde{r}^2 = (x_1/a_1)^2 + (x_2/a_2)^2 + (x_3/a_3)^2$, $(x_1,x_2,x_3)$ are the spatial coordinates, and $\{a_1,a_2,a_3\}=\{52.20, 45.44, 1153.2\}\,\mu$m.  Define also the trap Fermi energy (for a harmonic trap)

\begin{equation} E_F(N_t)= \hbar (\omega_1\omega_2\omega_3)^{1/3}(3 N_t)^{1/3},\label{2060}\end{equation}

\noindent used to scale the experimental energy data.  The two transverse and the axial trap frequencies are $\{\omega_1, \omega_2, \omega_3\}$ = $2\pi\{665,764,30.1\}$Hz, respectively, with $\omega_i=\sqrt{4U_0/ma_i^2}$ $(i=1,2,3)$, and $m=6.015$ amu is the mass of a $^6$Li atom.  The trap edge has $\mu_0=U(\vec{r})$, corresponding to

\begin{equation} \tilde{r} = \sqrt{\frac{1}{2}\, \mbox{ln}\left(\frac{U_0}{U_0-\mu_0}\right)}.\label{2070}\end{equation}

\noindent Clearly, $\tilde{r}$ increases as $\mu_0$ increases from $0$, and $\tilde{r}\to\infty$ as $\mu_0\to U_0$.

\par
An essential quantity in the data analysis of a function $y$ depending on $x$ is

\begin{equation} \chi^2\equiv \frac{1}{n}\sum\limits_i \left[\frac{(y-y_i)}{\sigma_i}\right]^2, \label{2080}\end{equation}

\noindent where $y$ and $y_i$ denote theoretical and experimental values, respectively, for the $i$'th of the $n$ data points, and $\sigma_i$ is the standard deviation for $y_i$.  If there are error bars on both the $x$ and the $y$ axes, we take \cite{Leo1987}

\begin{equation} \sigma_i^2\rightarrow\sigma_y^2+\left(\frac{dy}{dx}\right)^2 \sigma_x^2, \label{2090}\end{equation}

\noindent where $\sigma_x$ and $\sigma_y$ are the errors in $x$ and $y$, respectively.

\par
The Duke experiment measured $(E_t, S_t, N_t)$ directly, with the experiment done at constant $N_t\simeq1.3\times 10^5$.  There was no use of a heat bath or an atom bath, so $(T,\mu_0)$ for any particular data run were not known a priori.  I determine $(T,\mu_0)$ as needed by data fitting in the context of some theoretical LDA.  To connect some theoretical LDA to experimental data spanning a range of $T$, I proceed as follows: 1) Set some small $T$.  2) Set some $\mu_0$, and adjust its value until $N_t$ in Eq. (\ref{1990}) matches the experimental value.  3) Integrate over the trap with these values of $(T,\mu_0)$ to find the theoretical $E_t$ and $S_t$.  This step requires the transformation method $\{T, \mu_0, U(\vec{r})\}\to\{E_t, S_t, N_t\}$ described above in this Appendix.  4) Increment $T$ to a higher value and repeat with step 2 until we have a theoretical curve of $E_t /E_F(N_t)N_t$ versus $S_t/ k_B N_t$ spanning the full experimental data curve.\footnote{The division of the theoretical $E_t$ by $E_F(N_t)$ is done to match the experimental data, which is scaled this way.}  5) Calculate $\chi^2$ for the data consisting of $n$ pairs of $(S_t/k_B N_t, E_t / E_F(N_t) N_t)$.  6) Repeat this entire procedure with incremented $y_0$ and $y_1$ to minimize $\chi^2$ for the best fit between experiment and theory.

\par
\begin{figure}[tpg]
\centering
\includegraphics[width=4.0in]{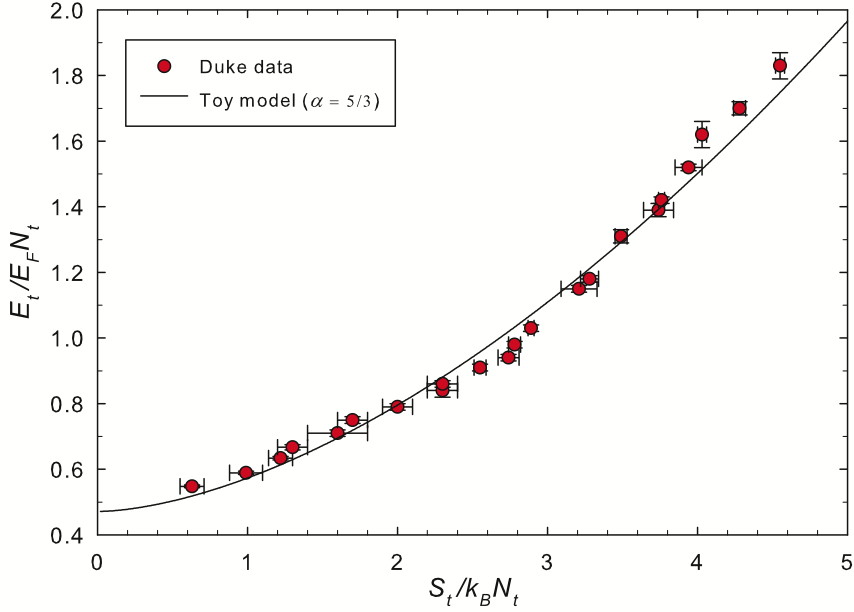}
\caption{The Duke data \cite{Luo2009} ($E_{840}/E_F$ versus $S_{1200}/k_B$) and the trap integrated toy model $Y(z)=0.2401 + 0.1392 z^{5/3}$, which was found to minimize $\chi^2=3.63$.}
\label{fig:7}
\end{figure}

Results are shown in Figure \ref{fig:7} for the toy LDA in Eq. (\ref{2005}), with the best fit on varying the two parameters $y_0$ and $y_1$ having $\chi^2=3.63$.  Clearly, this toy model, with just two fit parameters and no phase transition, does not produce a particularly good fit.  The results in section 4, with the two-piece LDA constructed from the geometric equation, are much superior.

\par
The Duke experiment operated at constant $N_t$, which in the context of the toy LDA Eq. (\ref{2005}) translates to constant $\mu_0$.  With the best fit in Fig. \ref{fig:4}, we have $\mu_0/U_0=0.0546$, a value which clearly has the atoms down near the bottom of the trap.

\section{APPENDIX 3: THE VIRIAL THEOREM}

In this Appendix I discuss the viral theorem

\begin{equation} E_t = 2 \int \rho(\vec{r})U(\vec{r})\,d^3 r, \label{3000}\end{equation}

\noindent which enables experimentalists to determine $E_t$ just by measuring density profiles $\rho(\vec{r})$.  Thomas {\it et al.} \cite{Thomas2005} argued that the virial theorem, valid for the ideal gas, holds as well in the strongly interacting Fermi fluid.

\par
However, the Thomas derivation \cite{Thomas2005} assumes implicitly that the pressure goes to zero at the edge of the trap.  This should be the case if the gas behaves like an ideal gas near the trap edge.  But with a power law divergence, a surface term involving the pressure enters the picture, as I will now demonstrate.  The Gibbs-Duhem equation at constant $T$ yields $\rho\,d\mu=dp$.  Since $d\mu=-dU$, we get

\begin{equation} \nabla p(\vec{r}) +  \rho(\vec{r})\nabla U(\vec{r}) = 0, \label{3010}\end{equation}

\noindent the condition of hydrostatic equilibrium.  Assume now a harmonic potential, for which $\vec{r}\cdot\nabla U(\vec{r}) = 2 U(\vec{r})$, and assume $p=2E/3V$, valid for the scaled fundamental equation in Eq. (\ref{60}).  Taking the dot product of both sides of Eq. (\ref{3010}) with $\vec{r}$, and integrating over the trapped sample, leads to

\begin{equation}  E_t = 2 \int\rho(\vec{r})U(\vec{r})\,d^3 r  + \frac{1}{2}\oint p(\vec{r})\, \vec{r} \cdot \hat{n}dA, \label{3020}\end{equation}

\noindent where $\hat{n}$ is a unit normal to the surface $\rho=0$.

\par
The surface term in Eq. (\ref{3020}) is zero if we assume that the thermodynamics at the trap edge is that of the ideal gas, resulting in the virial theorem Eq. (\ref{3000}).  For the monatomic ideal gas, we have the Sackur-Tetrode equation

\begin{equation} \displaystyle\frac{E}{N\epsilon_F} = e_0\,  \mbox{exp}\left(\frac{2S}{3k_B N}\right),  \label{3030}\end{equation}

\noindent where $e_0$ is a constant, and $\epsilon_F\propto\rho^{2/3}$.  Since $E=3N k_B T/2$, then for given $T$ as $\rho\rightarrow 0$, we get $z\rightarrow\infty$.  Also, $p=\rho k_B T$, which leads immediately to $p\rightarrow 0$ as $\rho\rightarrow 0$, and the surface term in Eq. (\ref{3020}) is zero.

\par
For the power law behavior, the series for $Y_H(z)$ Eq. (\ref{150}) yields a pressure

\begin{equation} p = \displaystyle \frac{4 \sqrt{\frac{6}{5}} m^{3/2}}{25 \pi ^2 \hbar ^3}\frac{(k_B T)^{5/2}}{ y_{-1}^{3/2}}, \label{3040} \end{equation}

\noindent which does not go to zero at fixed $T$ as either $\rho\rightarrow 0$ or $z\rightarrow\infty$.  The surface term in Eq. (\ref{3020}) will thus modify the virial theorem except at very small $T$.  But, this should not affect the analysis in this paper.

\end{document}